\documentclass[12pt]{article}
\usepackage{graphicx}
\usepackage{subfigure}
\usepackage[dvips]{color}
\usepackage{amssymb}

\input epsf.tex

\newcommand{\beq}{\begin{equation}}
\newcommand{\eeq}{\end{equation}}
\newcommand{\be}{\begin{equation}}
\newcommand{\ee}{\end{equation}}
\newcommand{\bea}{\begin{eqnarray}}
\newcommand{\eea}{\end{eqnarray}}

\makeatletter

\@addtoreset{equation}{section}
\makeatother

\def\href#1#2{#2}

\begin{document}

\baselineskip=15.5pt
\pagestyle{plain}
\setcounter{page}{1}

\begin{titlepage}
\begin{flushleft}
       \hfill                       FIT HE - 12-03 \\
       \hfill                       KYUSHU-HET 135 \\
       \hfill                       SAGA-HE-275 \\
\end{flushleft}

\begin{center}
  {\huge Holographic cold nuclear matter   \\ 
   \vspace*{2mm}
as dilute instanton gas  \vspace*{2mm}
}
\end{center}

\begin{center}

\vspace*{2mm}
{\large Kazuo Ghoroku${}^{\dagger}$\footnote[1]{\tt gouroku@dontaku.fit.ac.jp},
Kouki Kubo${}^{\ddagger}$\footnote[2]{\tt kkubo@higgs.phys.kyushu-u.ac.jp},
Motoi Tachibana${}^{Q}$\footnote[3]{\tt motoi@cc.saga-u.ac.jp},
\\
Tomoki Taminato${}^{\ddagger}$\footnote[4]{\tt taminato@higgs.phys.kyushu-u.ac.jp},
and Fumihiko Toyoda${}^{\P}$\footnote[5]{\tt ftoyoda@fuk.kindai.ac.jp}
}\\

\vspace*{2mm}
{${}^{\dagger}$Fukuoka Institute of Technology, Wajiro, 
Higashi-ku} \\
{
Fukuoka 811-0295, Japan\\}
{
${}^{\ddagger}$Department of Physics, Kyushu University, Hakozaki,
Higashi-ku}\\
{
Fukuoka 812-8581, Japan\\}
{${}^{Q}$Department of Physics, Saga University, Saga 840-8502, Japan\\}
{
${}^{\P}$Faculty of Humanity-Oriented Science and
Engineering, Kinki University,\\ Iizuka 820-8555, Japan}

\vspace*{3mm}
\end{center}

\begin{center}
{\large Abstract}
\end{center}
We study cold nuclear matter based on the holographic gauge theory, 
where baryons are introduced as the instantons in the probe D8/$\overline{\rm D8}$ branes
according to the Sakai-Sugimoto model. Within a dilute gas approximation of instantons,
we {search} for the stable states via the variational 
method and fix the instanton size. 
We find the first order phase transition from the vacuum to the nuclear matter phase
as we increase the chemical potential. At the critical chemical potential, we could see
a jump {in} the baryon density from zero to a finite definite value. 
While the size of the baryon in the nuclear matter is rather small 
compared to the nucleus near the transition point,
where the charge density is also small, it increases with the baryon density.
Those behaviors obtained here are discussed by relating
{them} to the force between baryons.

\noindent

\vfill
\begin{flushleft}

\end{flushleft}
\end{titlepage}
\newpage

\vspace{1cm}

\section{Introduction}

It is difficult to study dense nuclear matter from the 4D
non-perturbative approach like lattice gauge theory due to the sign
problem, which appears when we introduce the chemical potential (See 
for example \cite{Schafer:2005ff,Stephanov:2007fk}).
Then it is challenging to make clear the properties of the nuclear matter 
from the viewpoint of holographic gauge theory. 

In this approach, baryons
can be introduced as solitons on the probe flavor branes
\cite{SS,HSSY,Hashimoto:2008zw,Hashimoto:2009ys}. 
For the simplicity, we consider two-flavor case, $N_f=2$. The soliton,
which carries unit baryon number, is given as the BPST
instanton solution of the $SU(N_f)$ YM theory in the flat 5D space-time
of the probe brane \cite{HSSY}.
The instanton configuration has a scale parameter
which could be identified with the baryon size.
In the case of flat space, the size is free. On the other hand, 
the geometry of the world-volume of the embedded  flavored
brane is deformed in the fifth direction. As a result, the size of the
instanton is fixed by taking into account of this deformation. Actually,
in \cite{HSSY}, this size of the baryon is determined by minimizing the
action of the probe D8/$\overline{\rm D8}$, which
is expressed as a sum of the Dirac-Born-Infeld (DBI) and the
Chern-Simons (CS) terms. 
The CS term is also necessary to introduce the baryon chemical potential
in this approach with the instantons,
because the instantons, i.e. baryons, and the $U(1)$ gauge field dual
to the baryon number current are connected via the CS term. 

In this context, several approaches to study the nuclear system
have been performed by introducing the chemical potential of baryons
\cite{Kim:2006gp,BLL,Kim:2007vd,RSRW,Chuang:2010ku,Kaplunovsky:2012gb,Seki:2012tt,deBoer:2012ij}.
In \cite{BLL}, the instantons are, however, used to give a 
delta-function type source at the bottom of the D8 brane configuration
only in the CS term. Furthermore,
D4 branes are added to cancel the singularity generated by the
delta-function {type of} source at this point, then the
V-shaped D8/$\overline{\rm D8}$-brane configuration is obtained instead
of the smooth U-shaped one. As a result, the authors of \cite{BLL} have observed
a gapless transition from the vacuum to the nuclear matter phase at zero
temperature. 
However this point is 
{different from}
other many kinds of theories (See for example \cite{Cohen2004}).

On the other hand, the authors of \cite{RSRW} have preserved the flavor
gauge fields in the DBI action as well as in the CS term, then they
could find the first order 
{phase}
transition at finite baryon density.
However, the configuration used for the flavored gauge fields is obscure, 
so that it is difficult to see the relation between it and the
well known instanton. Furthermore, the profile of D8 brane is not solved
in the form of U-shape one, which should depend on the
gauge fields configurations in the same DBI action. 

\par
Here we introduce the explicit form of
instanton solution, which is exact in the flat 4D space (our three-space
and fifth one), by keeping its size parameter. Then it has a smooth
configuration in the fifth coordinate direction instead of the delta
function form. Furthermore, the flavored Yang-Mills fields are retained
as instantons in the DBI action of the D8 probe brane 
in its leading order, namely up to the square of the field strength. The
latter point is crucial to find a gap of the baryon density at the
transition point as in the case of \cite{RSRW}. Differently from
\cite{RSRW}, in our approach, the flavored YM field is solved by
determining the remained size parameter of the instanton. 
After that, the physical quantities like chemical potential and baryon
number density are obtained to search for the phase transition. 

As for the profile of D8/$\overline{\rm D8}$ brane, it is restricted
here to the antipodal solution. Then,
we can use a simple D8 brane profile \cite{HSSY}, which is obtained
without any instantons as an antipodal U-shaped configuration.
In general, the lowest energy solution for some
finite  baryon density  $n$ is not equivalent to the simple solution
mentioned above, since it is not antipodal. However,
the solution is always approximated by the simple one with a negligible
correction for any $n$.
This implies that the simple solution is useful at any $n$ as the
D8/$\overline{\rm D8}$ antipodal configuration in our analysis. This is
the reason why we restrict 
the profile to the antipodal solution. Another reason is for the
simplicity of the present analysis.

Through our analysis, we find a first order phase transition, which is expressed
in $\bar{n}$-$\mu$ plane, where $\bar{n}$ and $\mu$ denote
the vacuum expectation value of charge density (times some dimensionful
constant) and the chemical potential.
At the transition point, the baryon density jumps from zero to a finite
value, which corresponds to the transition from the
vacuum to a nuclear matter phase. 
By adjusting the parameters, at
the transition point, we observe the baryon mass as
$\mu_{B}\sim 2.3{\rm GeV}$ which is rather large compared to the realistic
nucleon mass. Furthermore at this transition point, the size of the baryon is rather small, but it
increases as $\sqrt{\mu}$ at large $\mu$. 
 

\vspace{.3cm}
The outline of this paper is as follows. In the next section, the dilute instanton gas model is set up.
Then, the embedded solution of the D8 profile and other fields are given in the section 3. In the section 4, energy density of the system
is examined by using a simple profile solution to find phase transitions.
In the final section, summary and discussions are given.  

\section{Dilute instanton gas model}\label{sec21}

Here baryon is identified with an instanton \cite{HSSY}. Then
we approximate multi-baryon system by the dilute gas system of multi-instantons. The system is
described by using D8 branes with the following action.

\subsection{DBI action}

For stacked two branes,
\beq
  S_{D_p}=-T_p\int d^{p+1}\xi^a e^{-\Phi}{\rm Str}L\, ,
\eeq
where $p=7 (8)$ for type IIB(A), and ${\rm Str}$ denotes the symmetric trace of flavor $U(2)$
\bea
  L&=&\sqrt{-{\rm det}\left(f_0+\vec{f}_1\right)}\, , \\
  (f_0)_{ab}&=&\left(G_{MN}\partial_a X^M\partial_b X^N+B_{ab}+\tilde{F}_{ab}\right)\tau_0\, , \\
  B_{ab}&=&B_{MN}\partial_a X^M\partial_b X^N=-B_{ba}\, \\
  \tilde{F}_{ab}&=&2\pi\alpha'F_{ab}\, , \\
  \vec{f}_1&=&2\pi\alpha' F_{ab}^i\tau_i\, ,
\eea
where $a,b=0\sim p$, $M,N=0\sim 9$, and
 $\tau_0$, $\tau_i$ are the unit, Pauli's spin matrices. The Str part is expanded as
\bea
  {\rm Str}L&=&{\rm Str}\left\{\sqrt{-{\rm det}(f_0)}\left(1-{1\over 4}{\rm tr}x^2
            +{1\over 8}({\rm tr}x)^2+\cdots  \right)\right\}\, , \\
         x&=&f_0^{-1}\vec{f}_1\, ,
\eea
where '${\rm tr}$' denotes the trace of the coordinate index. The linear term vanishes for
${\rm Str}$, then it is dropped.
The coordinates are set as
$$ (\xi^0,\xi^1,\xi^2,\xi^3,\xi^4,\cdots)=(x^0,x^1,x^2,x^3,z,\cdots)$$
and we make the ansatz
\bea
   \tilde{F}_{ab}&=&2\partial_{[a}A_{b]}\, , \quad B_{ab}=0 \, ,\\
   A_b&=&A_b(z)\delta_b^0\, ,\\
   (\vec{f}_1)_{ij}&=&Q(x^m-a^m,\rho)\epsilon_{ijk}\tau^k\, ,\label{an1} \\
   (\vec{f}_1)_{iz}&=&Q(x^m-a^m,\rho)\tau^i\, ,\label{an2}
\eea
where $\rho$ ($a^m$) denotes the instanton size (position), $\epsilon_{123z}=1$, 
$i,j=1,2,3$ and  $m=1,\dots ,4$, where $x^4=z$.

\vspace{.3cm}
Then the above series of $\vec{f}_1$, by retaining its lowest order, are obtained as 
\bea
  {\rm Str}L&=&2\sqrt{-{\rm det}(f_0)}\left(1+{3\over 2}Q^2\left[\left({\cal G}^{11}\right)^2
        +{\cal G}^{11}{\cal G}^{zz}\right]\right)\,,  \label{Inst-Contri-1}\\
      {\cal G}_{ab}&=& G_{MN}\partial_a X^M\partial_b X^N \, . \label{Inst-Contri-2}
\eea
Hereafter we neglect the higher order terms of the non-Abelian gauge fields.

\vspace{.5cm}
\subsection{{$Q$} as a Dilute Gas of Instantons }

In order to see the baryon spectrum, $Q$ is given as an instanton solution in flat 4D space $\{x^m\}$
\cite{HSSY}, 
\beq\label{Inst-Sol}
   Q={2\rho^2\over \left((x^m-a^m)^2+\rho^2\right)^2}\, .
\eeq
However, it is not a solution of the equation of motion given by the D8 brane action. We use it as a trial function
which is supposed to be a solution of the system. This is confirmed in our formulation such that
the size
parameter $\rho$ of the given instanton configuration is determined to satisfy the variational principle of the
energy density up to the leading order of the expansion given in (\ref{Inst-Contri-1}).

According to the above strategy, we study multi-baryon state by replacing $Q(z)$
by the multi-instanton form with dilute gas approximation. Then we
rewrite $Q$ as
\beq\label{multi-i}
  Q^2=\sum_i^{N_I}{4\rho^4\over \left((x^m-a^m_i)^2+\rho^2\right)^4}\, ,
\eeq
where the overlapping between the instantons are suppressed in obtaining $Q^2$.
So we find the energy density in the flat 4D space as a sum of each single instanton,
\bea
  \int d^4\xi^m Q^2&=& 2N_I\int_0^{\infty} dz \bar{q}(z)^2\, \nonumber \\
  &=&{2\pi^2\over 3}N_I\, ,
\eea
where
\beq\label{multi-i-2}
  \bar{q}^2={\pi^2\rho^4\over 2(z^2+\rho^2)^{5/2}}\, 
\eeq
for the case of dilute gas, where interactions between instantons are neglected. Then the result
is given by one instanton ``mass'' times their number $N_I$. 

As for the size parameter $\rho$, we determine it by minimizing the embedded D8 brane energy
as in \cite{HSSY}. We notice that there is another method giving $Q$ including its $z$ dependence by solving 
the embedding equations of motion \cite{RSRW}. We, however, solve only for the size parameter here.

\subsection{CS term} 
Supposing a reduced 5D action, we consider the following form of CS term for $N_f=2$ \cite{HSSY},
\beq
  S_{CS}={N_c\over 24\pi^2}\epsilon^{m_1\cdots m_4} \int
  d^4xdz{3\over 4} A_0{\rm Tr}(F_{m_1m_2}F_{m_3m_4})\, .
\eeq
For the instanton configuration, we have
\beq
  S_{CS}={3N_c\over 2\pi^2}\int d^4xdz A_0 {Q^2\over
  (2\pi\alpha')^2} \, .
\eeq
This term is then included in our calculation as the coupling of $A_0$ and instantons
by supposing $Q=Q(z)$ as in the DBI term.

\section{Solution of embedded D8/$\overline{{\rm {\bf D8}}}$ brane}

Induced metric for D8 brane is given as \cite{SS,HSSY},
\beq\label{induced-metric-d8}
 ds_9^2={\lambda l_s^2\over 3}\left({4\over 9}k^{1/2}(z)\eta_{\mu\nu}dx^{\mu}dx^{\nu}
 +g(\tau){4\over 9}k^{-5/6}(z)dz^2+k^{1/6}(z)d\Omega_4^2\right)\, ,
\eeq
\beq
 k(z)=1+z^2\, , \quad e^{-\Phi}={3^{3/2}\pi N_c\over\lambda^{3/2}}
 k(z)^{-1/4}\, ,
\eeq
where $k\equiv U^3=U_{KK}^3+U_{KK}z^2$ and we set 
\beq
U_{KK}=1, \quad R^3=\frac{9}{4}=\pi g_sl_s^3N_c,\quad
\lambda=3\sqrt{\frac{\pi g_sN_c}{l_s}}\, , \quad M_{KK}={3U_{KK}^{1/2}\over 2R^{3/2}}\, ,
\eeq
where $l_s$ and $g_s$ are the string length and the string coupling
respectively \footnote{We notice that $z$ is taken as a world-volume coordinate of the
D8 brane, and $\tau$ is set as the external one. Then it is included in the induced metric 
of D8 brane as a scalar field $\tau(z)$.}
In this notation, we use $\lambda l_s^2/3=3/2$,
$\alpha'=9/2\lambda$.
Moreover, we give
\beq
  g(\tau)=1+z^2k^{1/3}\tau'^2\, , \quad \tau'={\partial\tau(z)\over \partial z}\, .
\eeq
Note that $\tau$ should have a periodicity $\tau\sim \tau+\delta \tau$
in order to avoid a conical singularity at $z=0$. In the case of
(\ref{induced-metric-d8}), $\delta \tau$ is denoted as $2\pi$ \cite{SS}.

Then we have 
\bea
  S_{D8}&=&-T_p\int d^{9}\xi^a e^{-\Phi}{\rm Str}L\, , \nonumber \\
  &=&-{N_c\lambda^3\over 3^9\pi^5}\int d^4x dz k^{{1\over12}}(z){\rm Str}
   \sqrt{-\det (g_{MN}^{(5)}+\tilde{F}_{MN})}\, , \\
  g_{MN}^{(5)}dx^Mdx^N&=&k^{1/2}(z)\eta_{\mu\nu}dx^{\mu}dx^{\nu}+ g (\tau ) k^{-5/6}(z)dz^2\, , \\ 
  \tilde{F}_{MN}&\equiv&\frac{3}{2}(2\pi\alpha')F_{MN}={27\pi\over 2\lambda}F_{MN}\, .
\eea
Using the above formula, we find
\beq
   S_{D8}=-2\kappa\int d^4x dz k^{5/6}(z)\sqrt{g(\tau)k^{-1/3}(z)-\tilde{E}_z^2}
   \left(1+{3\over 2}\tilde{Q}^2\left(k^{-1}+k^{1/3}g(\tau)^{-1}
   \right)\right)\, ,
\eeq
where $\tilde{E}_z=\tilde{F}_{z0}=\partial_z\tilde{A}_0(z)$ and $\tilde{Q}=\frac{3}{2}Q$. Similarly we obtain
\beq
 S_{CS}={2^2\lambda^3N_c\over 3^8\pi^5}\int d^4x dz \tilde{A}_0(z)\tilde{Q}^2\, .
\eeq
Hereafter we drop all tildes for simplicity.

The effective action is given as
\bea\label{multi-1}
  S&=&S_{D8}+S_{CS}=-2\kappa V_3\int dt dz L_Q\, , \\
  L_Q&=&k^{5/6}\sqrt{g(\tau)k^{-1/3}-E_z^2}~Q_1 -{6}A_0 n\bar{q}_0^2\, ,  \\
    Q_1&=&1+{3\over 2}n\bar{q}_0^2\left( k^{-1}+k^{1/3}g(\tau)^{-1}
    \right)\, ,
\eea
where $V_3$ is the volume of 3-dimensional space and $n$ is the density of instantons defined by
\beq
    n={N_I\over V_3}\, .
\eeq
Note here that
we replaced the previous instanton configuration $ \bar{q}^2$ to the following new form
\beq\label{multi-i-3}
  \bar{q}_0^2={9\over 8}{\pi^2\rho^4\over \left((z-z_0)^2+\rho^2\right)^{5/2}}\, ,
\eeq
which is reduced to $ \bar{q}^2$ for $z_0=0$. In any case, we take the position of the inserted instantons at the bottom
of the D8 brane since they are pulled to the horizon by the gravitational force. In other words, we put the instantons
at their stable position. On the other hand, they are distributed at random points 
in the three space of our Minkowski space-time.

\subsection{Equations of motion for the system $S=S_{D8}+S_{CS}$}

We suppose the form (\ref{multi-i-3}) for the instanton contribution, where the parameter $\rho$ is remained
to be determined. It is given here by minimizing the action. 

\vspace{.3cm}
The equations of motion for  $A_0(z)$ and $\tau(z)$ are given as
\beq\label{eq-A0}
  \partial_z\left(k^{5/6}{E_z\over \sqrt{g(\tau) k^{-1/3}-E_z^2}}Q_1\right)= 6n \bar{q}_0^2\, ,
\eeq
\beq\label{eq-tau}
   \partial_z\left(k^{5/6}{z^2\tau'\over \sqrt{g(\tau) k^{-1/3}-E_z^2}}Q_1\right)
   -{3n}\partial_z\left(k^{3/2} \sqrt{g(\tau) k^{-1/3}-E_z^2}{ \bar{q}_0^2z^2\tau'\over g^2(\tau)}\right)=0\, .
\eeq

At first, we solve for $A_0(z)$, whose equation of motion is given as
\beq\label{E}
   k^{5/6}{E_z\over \sqrt{g(\tau) k^{-1/3}-E_z^2}}\left(1+{3\over
   2}n\bar{q}^2\left(k^{-1}+k^{1/3}/g(\tau) \right)\right)=d\, , 
\eeq
\beq\label{d}
    d={6}n\int_{z_0}^{z}~dz \bar{q}_0^2+C={9\over 4}{n\pi^2}(z-z_0){2(z-z_0)^2+3\rho^2\over ((z-z_0)^2+\rho^2)^{3/2}} +C\, . 
\eeq
{Here we take the integration constant $C$ as zero since the displacement would vanish when instantons disappears
(or $n=0$).} Then we have
\beq
   E_z={g^{1/2}k^{-1/6}\over \sqrt{1+f^2}}\, , \quad f={k^{5/6}\over d}Q_1\, .
\eeq

In the next, by substituting the above $E_z$ into (\ref{eq-tau}), we find
\beq\label{eq-tau-2}
 {z^2k\tau'  \over g^{1/2}(\tau) } \left\{{\sqrt{1+f^2}\over f}Q_1
   -{f\over \sqrt{1+f^2}}{{3n}k^{1/3}  \bar{q}_0^2\over g(\tau)}\right\}=c_0\, ,
\eeq
where $c_0$ denotes an integration constant, which is related to the bottom coordinate $z_0$. The solution $\tau(z)$
has in general has a U-shape configuration in the $\tau$-$z$ plane. And, we find $\tau'(z_0)=\infty$ at its bottom point, then
the following relation is obtained from Eq.(\ref{eq-tau-2})
\beq\label{eq-z0}
  z_0Q_1(z_0)k^{3/4}(z_0)=c_0\, .
\eeq

\subsection{Simple solution in the antipodal case}
{Here we consider the antipodal solution which satisfies $\tau(\infty)=\pi/2$ (or $-\pi/2$). 
The explicit and exact (U-shaped) form of this solution is given as the solution of Eq.(\ref{eq-tau-2}) at $n=0$ 
and in the limit of $z_0\to 0$. 
It is approximately given as
\beq\label{simple-tau-sol}
\tau'(z)=0\quad {\rm for}\quad z>z_0.
\eeq
In general, the configuration of the solution $\tau(z)$ depends on the instanton
density $n$, and $\tau(\infty)$ increases with $n$ when $z_0$ is kept small.
Therefore we consider here a new type of solution in order to obtain the
antipodal solution in the limit of $z_0\to 0$. 
In this case, all such solutions are approximated here by a simple solution of Eq.(\ref{eq-tau-2}).
\footnote{The details of the solutions and justification of this approximation are given in the Appendix~\ref{app:Simple-sol}.}
}

On the other hand, we can consider other antipodal solutions that satisfy $\tau(\infty)=\pi/2$ at finite $z_0$ and $n$. However, it is confirmed by numerical analyses that the energy of the solution at finite $z_0$ is always larger than the solution (\ref{simple-tau-sol}) in the same $n$. 
Therefore, we adopt here
the solution (\ref{simple-tau-sol}).

Then $E_z$ and the chemical potential are obtained by noticing $g(\tau)=1$ as
\bea
  E_z&=& { k^{-1/6} d\over \sqrt{Q_0^2k^{5/3}+d^2}}\, , \label{Ez} \\
  Q_0&=&1+{3\over 2}n{\bar{q}_0^2}(k^{-1}+k^{1/3})\, ,  \label{Q1} \\
  \mu &=&\mu_c+\int_0^{\infty}dz E_z\, .  \label{chemical-pot} \label{mu-s}
\eea
Here the quark chemical potential \footnote{We notice here that the chemical potential
given above expresses the one of the quark rather than that of the baryon.} $\mu$ and $\mu_c$ are defined as
\beq
  \mu=A_0(\infty)\, , \quad \mu_c=A_0(0)\, .
\eeq

In this case,  $E_z$ is expanded at large $z$ as follows
\beq
  E_z={9n\pi^2\over 2}\left({1\over z^{2}}-{1\over z^{4}}+\cdots \right)\, ,
\eeq
{Here we notice the coefficients of the leading and sub-leading orders are independent of $\rho$.}
Then we obtain
\beq
  A_0=\mu-{9n\pi^2\over 2}{1\over z}+\cdots\, .
\eeq
The charge density is expected to correspond to the coefficient of the
sub-leading term as
\beq
 c \langle\Psi^{\dagger}\Psi\rangle=\bar{n}={9n\pi^2\over 2}\, ,
\eeq
where $c$ is a dimensionful coefficient obtained by returning dimensionless variables to
dimensionful ones.

In the small $z$ region,  $E_z$ is expanded as 
\beq
  E_z={9n\pi^2\over 2}\left({a_1z}+{a_2 z^{3}}+\cdots \right)\, ,
\eeq
where $a_{1,2}$ are the functions of $\rho$. Then we find the asymptotic
solution of $A_0$,
\beq
   A_0=\mu_c+{9n\pi^2a_1\over 4}z^2+\cdots \, .
\eeq
where $\mu_c$ is undetermined.

\vspace{.5cm}
\section{Energy density and phase transition}

The Chern-Simons term is rewritten as
\bea
-{6n} \int_0^{\infty}dz A_0 \bar{q}_0^2&=& 
    -{6n} \int_0^{\infty}dz A_0 ( \partial_z q_b)\, \nonumber\\
    &=& {6n}\left( \int_0^{\infty}dz E_z q_b-{3\pi^2\over 4}\mu\right)\,\nonumber \\
     &=& {6n}\left( \int_0^{\infty}dz E_z \left(q_b-{3\pi^2\over 4}\right)-{3\pi^2\over 4}\mu_c\right)\, ,
\eea
where 
\beq
   q_b={3\pi^2\over 8}{2(z-z_0)^3+3(z-z_0)\rho^2\over \left((z-z_0)^2+\rho^2\right)^{3/2}}\,,
\eeq
and we used $q_b(\infty)=\pi^2/3$, $q_b(z_0)=0$ and Eq.(\ref{chemical-pot}). We notice about the sign of the
$S_{CS}$ term. When it is taken as minus as given above, and then we use
the self-dual instanton solution. In this case we find positive
chemical potential as shown in Eq.(\ref{chemical-pot}). On the other hand, for opposite sign and self-dual solution,
we obtain opposite sign chemical potential
\beq
 \mu =\mu_c-\int_0^{\infty}dz E_z\, .  \label{chemical-pot-2}
\eeq
and the same sign energy density except for the chemical potential part,
\beq
   +{6n} \int_0^{\infty}dz A_0 \bar{q}_0^2=
   {6n}\left( \int_0^{\infty}dz E_z \left(q_b-{\pi^2\over 3}\right)+{\pi^2\over 3}\mu_c\right)\, .
\eeq

When we use anti-self dual solution, then the same expression is obtained for the opposite sign. So we must use
the sign in our case, we should use the self-dual solution (baryon). When we apply the same analysis for 
the anti-baryon,
we must change the sign of $S_{CS}$ and use the anti-self dual solution. Then we find the same result in both cases.

\vspace{.5cm}
By using the above solution for $A_0$, we define the energy density as
\beq
  E(n)=V_3\left(\int dzL_Q-\bar{n}\mu_c \right)\, ,
\eeq
where $L_Q$ is given  as
\bea
  L_Q&=&L_{DBI}+L_{CS}\, , \\
  L_{DBI}&=&{k^{2/3}Q_0^2\over \sqrt{Q_0^2+k^{-5/3}d^2}}\, , \\
  L_{CS}&=&{6n}d{q_b-{3\pi^2\over 4} \over
         k \sqrt{Q_0^2+k^{-5/3}d^2}}\,.
\eea

Then, we compare the energy density $E(n)$ with $E(0)$ to find the critical value of chemical potential, $\mu_{cr}$.
Above which we will find $E(n)<E(0)$ then the baryon phase is realized there.

 The integration of
the Lagrangian $L_{DBI}$ for finite $n$ diverges, so we need a subtraction, which can be chosen as the energy
density of the vacuum with $n=0$.
$E(0)$ is given as
\beq
    E(0)=V_3 \int dz k^{2/3}\, .
\eeq
Then we can perform the calculation of $E(n)-E(0)$ by using
$\tilde{L}_{DBI}$, $$\tilde{L}_{DBI}=L_{DBI}-k^{2/3}. $$ 
We can thus estimate the energy difference for a given density $n$ by normalizing as
\beq\label{energy-d}
  {\cal E}=\int_0^{\infty} dz\left[\tilde{L}_{DBI}+L_{CS}\right]-\bar{n}\mu_c\, .
\eeq
In estimating this energy density, we set $\mu_c=A_0(0)=0$ since $A_0(z)$ must be a two valued function which
is symmetric at $z=0$. Performing the integration over $z$ for a fixed $n$, we obtain $ {\cal E}(\rho)$
as a function of $\rho$. 

In the estimation of $ {\cal E}(\rho)$, we must be careful about the value of $ {\cal E}(0)$. 
The value is sensitive to the order of taking the limit of $z\to z_0$ and $\rho\to 0$. Our estimation
is performed by the integration with respect to $z$ from $z_0$ to $z=\infty$, then the limit of $\rho\to 0$
is examined.

\vspace{.3cm}
\subsection{Phase Transition}

\begin{figure}[htbp]
\begin{center}
\includegraphics[width=7cm]{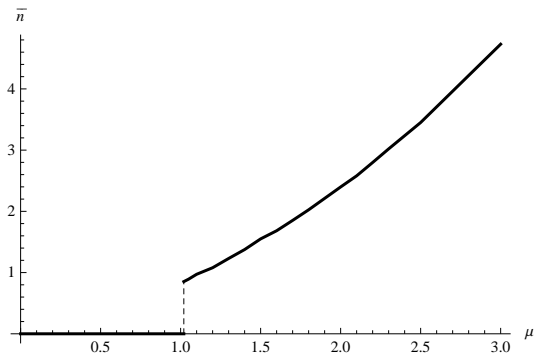}
\caption{{\small The diagram of
 $(\mu, \bar{n}=\frac{9\pi^2}{2}n)$ for the simple solution. The first order phase transitions, from vacuum to the
 nuclear matter phase at $\mu_{cr}=1.0176$, is seen.
}}
\label{critical-1}
\end{center}
\end{figure}

{To estimate physical quantity, we must determine the  
size parameter $\rho$ by minimizing the energy density ${\cal E}(\rho)$.
This is performed by estimating ${\cal E}(\rho)$ for each fixed $\mu$ not for fixed $n$
since the parameter which characterizes 
the boundary theory is not $n$ but the chemical potential $\mu$. 
In our formalism, however, the energy density is given as a function of  $n$ and $\rho$ from
the Eq. (\ref{energy-d}). At the same time, $\mu$ is also
obtained as a function of $n$ and $\rho$ from Eq. (\ref{chemical-pot}). Then ${\cal E}(\rho,\mu)$ for a fixed
$\mu$ is estimated by using both calculations of ${\cal E}(\rho,n)$ and $\mu(\rho,n)$ although it is a little
complicated.

\begin{figure}[htbp]
\begin{center}
\includegraphics[width=7.0cm]{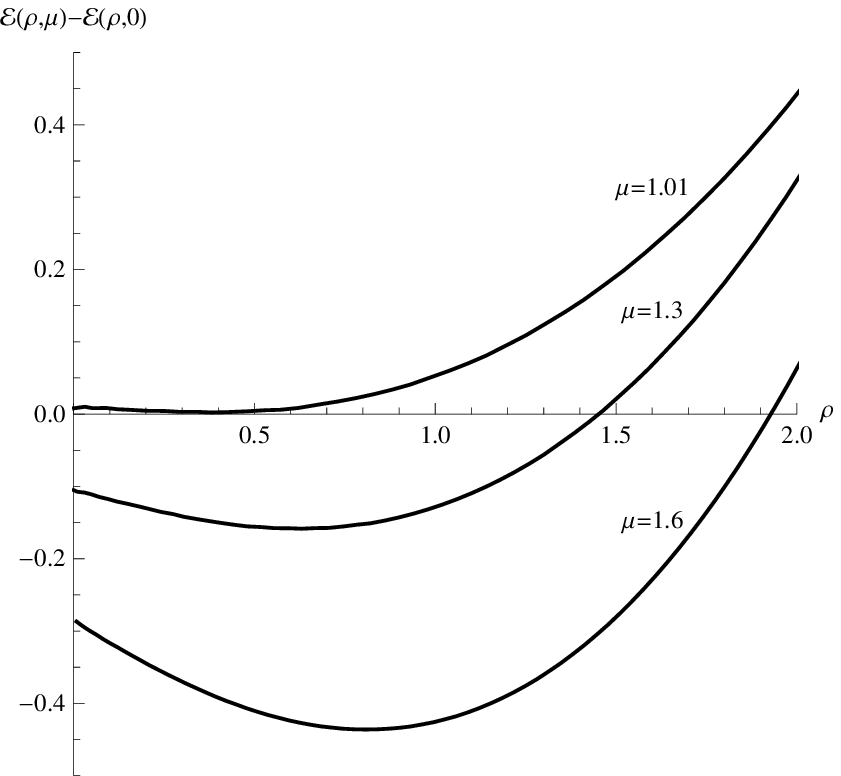}
\includegraphics[width=7.0cm]{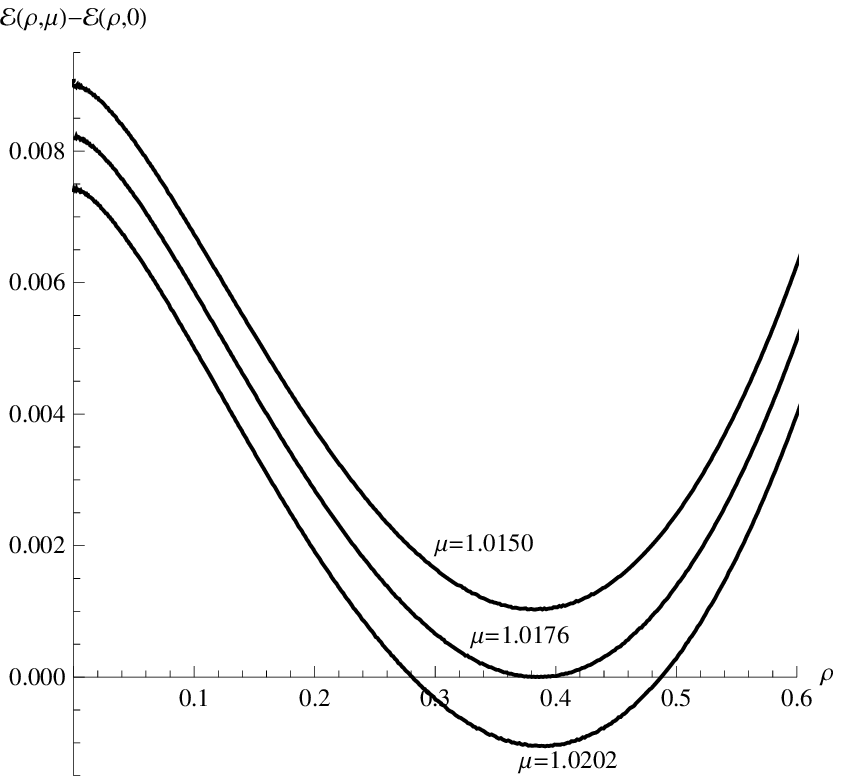}
\caption{{\small  $ {\cal E}(\mu)- {\cal E}(0)$ versus instanton size
 $\rho$ near phase transition point $\mu_{cr}\sim 1$.
}}\label{delta-E2}
\end{center}
\end{figure}

Noticing the property of ${\cal E}(\rho)$ mentioned above, we estimated ${\cal E}(\rho)$ for 
a fixed $\mu$. 
Then, fix the the value of $\rho$ as $\rho_m$ at the minimum of energy density, namely as 
${\cal E}_{\rm min}={\cal E}(\rho_m)$ for the given $\mu$.
After that, we determine $\bar{n}$. The resultant curves of ${\cal E}(\rho)$ are shown in the 
Fig.  \ref{delta-E2} near the phase transition point. The phase transition is observed at $\mu_{cr}=1.0176$, where
${\cal E}(\rho_m)$ changes its sign from positive to negative. This implies the transition from the vacuum to the 
nuclear matter phase at a finite charge density and finite chemical
potential, $(\mu,\bar{n})=(1.01,0.85)$. This transition is also seen more clearly from 
the Fig. \ref{critical-1} represented in the plane of $\mu$-$\bar{n}$.
}

The reason why $\bar{n}$ is finite at the transition point is reduced to the preserved flavored
gauge fields in the DBI action, this is reflected to the factor $Q_0>1$ given in (\ref{Q1}).
This point is also mentioned in the introduction. In our model, this is explicitly understood as
follows. When we set $Q_0=1$, the integrand in (\ref{energy-d}) is expanded near $n=0$ as,
\beq\label{n-exp}
  \tilde{L}_{DBI}+L_{CS}=6nd{q_b-3\pi^3/4\over k}-{1\over 2k}d^2+\cdots\, ,
\eeq
where $\cdots$ denotes the higher order terms of $n$, and notice $d$ is the order of $n$.
It is easy to see that the first term of the right hand side of (\ref{n-exp}) is non-positive since
\beq
  q_b-{3\pi^3\over 4}\leq 0\, .
\eeq
Here the equality is obtained for $\rho=0$. Then the energy density is always negative for $n>0$
in this case, namely for $Q_0=1$. This implies the transition occurs at $n=0$ when we neglect
flavored gauge fields in the DBI action. So we can say that the interaction among the baryons 
as dilute gas is 
included in the model
through the preservation of the flavor gauge field configuration in the probe brane.
In spite of the dilute gas approximation adopted, the attractive force to form the nuclear matter phase
would be reduced to the curved space in the fifth coordinate direction, namely the interaction
reduced to the vacuum of the Yang-Mills theory. Although the details of the dynamics should be examined more,
the importance of the flavored gauge fields in DBI is clearly understood from our model setting.

\vspace{.3cm}
\noindent{\bf Size of baryon}

\begin{figure}
\begin{center}
\includegraphics[width=7.0cm,height=6cm]{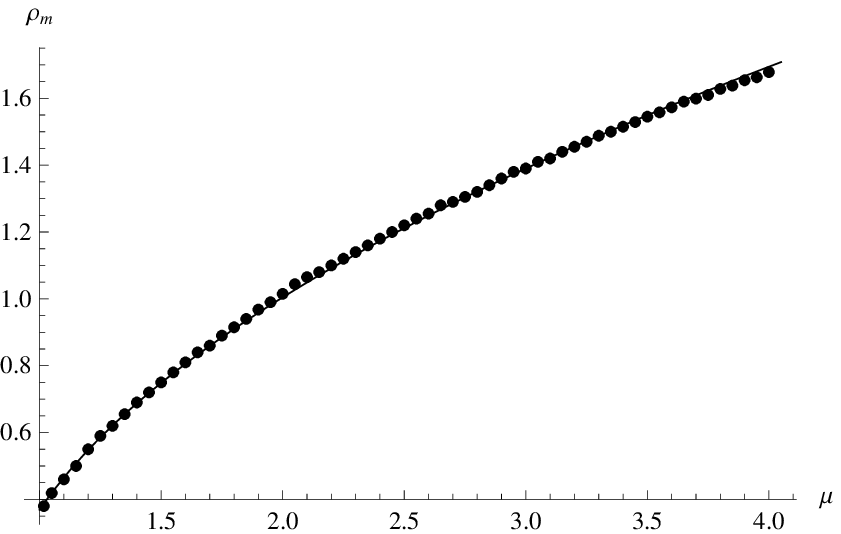}
\includegraphics[width=7cm,height=6cm]{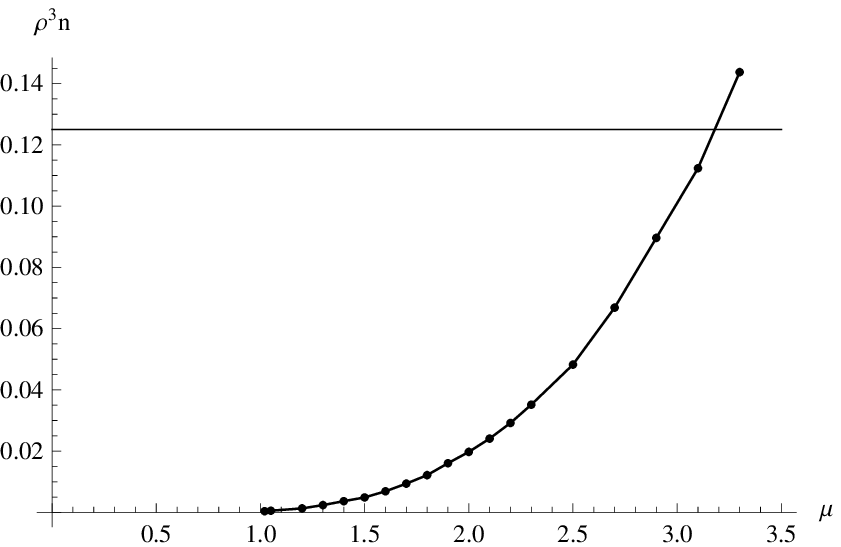}
\caption{{\small 
(Left) The Plot of ($\mu$, $\rho_m$). The solid line is given as
 $\rho_m=\sqrt{\mu-0.78}-0.10$.
(Right) The relation $\mu$-$\rho^3n$ for $\lambda=16.6$. The horizontal
 line at $\rho^3n=1/8$ corresponds to a rough accommodation limit of the
 dilute gas approximation. The intersection point at
 about $\mu\sim 3.1$ denotes the useful limit of the
  model.
}}
\label{critical-2}
\end{center}
\end{figure}

{As for the size of the baryon, which is identified with $\rho_m$, its $\mu$ dependence
is shown in the left of the Fig. \ref{critical-2}. From this numerical results, we
can see that $\rho_m$ grows with $\mu$ monotonically, and it is fitted by the curve
$$\rho_m\simeq\sqrt{\mu-0.78}-0.10\, .$$
This implies $\rho_m$ increases with the increasing baryon
density since $\mu$ grows with $n$.
This behavior could be naturally understood by
considering the attractive force between the baryons to form the stable nuclear matter.
This force becomes large with increasing $n$ since 
the number of the nearest baryons of one baryon
grows. As a result, the attractive force between them increases, and then the size of the baryon
is stretched to larger size.

However, at some critical density, baryons begin to overlap each other, then it would become difficult
to discriminate each baryon spatially. In this case, we will find a new nuclear phase, but our model
cannot be applied to this density since the dilute gas approximation is
no longer useful in this region.}
{In the right of the Fig. \ref{critical-2}, this point is seen
by considering a cube surrounding instanton in three dimensional
space which the length is equal to $2\rho$. 
We can see from {this figure that  
the dilute gas
approximation is not valid any more for $\mu>\mu_{{\rm limit}}$.} 
We notice here the following point related to the parameter $\lambda$. Since
the critical value of $\mu$ 
depends on $\lambda$ when we adjust the parameters, we find that the point
$\mu_{{\rm limit}}$ increases as $\lambda$ deceases.
{In the case of  $\lambda>155$, our model would be no longer valid since $\mu_{{\rm limit}}$ become smaller
than $\mu_{{\rm cr}}$. }
}

\vspace{.3cm}
\noindent{\bf Numerical estimation of the size}

{In the above analysis, the physical quantities are set to be dimensionless. 
In this setting, we get
\beq
\mu_{cr}\simeq 1.01\, , \quad
n_{cr}\simeq 0.019\  (\bar{n}_{cr}\simeq 0.85)\, , \quad \rho_m\simeq 0.38
\eeq
at the critical point.
So we must 
see their dimensionful values to compare them with the realistic
physical values, ($\mu_B=N_c~\mu$, $n_B$, $\rho_B$). 

In our notation, furthermore, we have set as\footnote{
In reality, we dropped the factor $2\pi\alpha^\prime$ in
(\ref{Inst-Sol}). However this factor can be absorbed into $n$ because they
always arise together. So we can recover this factor by interpreting $n$
as $(2\pi\alpha^\prime)^2n$.}

\beq
\frac{1}{\sqrt{2N_f}}\frac{27\pi}{2\lambda}A_0\to A_0,\quad (2\pi\alpha')^2n=\left(\frac{9\pi}{\lambda}\right)^2n\to n
\eeq
Therefore when we compare $\mu_{cr}$, ${n}_{cr}$, and $\rho_m$ given above with physical
quantities of baryon, their scales are given by $M_{KK}$ and $\lambda$ from the
following relations as
\bea
\frac{\mu_B}{M_{KK}}\!\!&=&\!\!{\sqrt{2N_f}N_c}\mu_{cr}\frac{2\lambda}{27\pi} \simeq {6}\times1.01\times\frac{2\lambda}{27\pi},\label{note1} \\
 \frac{n_B}{M_{KK}^3}\!\!&=&\!\!n_{cr}\frac{1}{(2\pi\alpha')^2} \simeq 0.019\times\frac{1}{(2\pi\alpha')^2}
=0.019\left(\frac{\lambda}{9\pi}\right)^2\!\!,\label{note2}
\\
\rho_B M_{KK}\!\!&=&\!\!\rho_m \simeq 0.38.\label{note3}
\eea
Here we used $N_f=2$ and $N_c=3$.

In the Sakai-Sugimoto model\cite{SS,Hashimoto:2008zw}, these parameters are chosen as $M_{KK}\simeq 0.949$ GeV and $\lambda=16.6$ to fit the values of $\rho$ meson mass and the pion decay constant
in the vacuum. Then these parameters lead to 
\beq
\mu_B\simeq {2.3} \,\mbox{GeV},\quad n_B\simeq
{0.70}\, \mbox{fm}^{-3},\quad
\rho_B\simeq 0.079\, \mbox{fm}.
\eeq

}

This value of $n_B$ is close to $4n_0$ where $n_0=0.16 fm^{-3}$ is the density of the normal nuclear matter.
It is interesting that $4n_0$ corresponds to the density of the inner core($n>2n_0$) of a neutron star  \cite{HPY}.

\section{Summary and Discussions}

We have studied cold nuclear matter 
using the Sakai-Sugimoto model, 
{where the baryon is introduced as a soliton on the probe flavor branes.
The soliton, which carries unit baryon number, is given by the BPST instanton solution
in the probe D8 branes.} Multi-baryon system is approximated by the dilute gas of instantons 
in the brane action. The profile of the D8 brane depends on the configuration of the 
instanton and its density. So it is rather complicated to solve them from the given D8 brane action.

The instanton configuration used here has the same form
as 
the one given in the flat
4D space, but it is not the solution
in the deformed bulk background. So we recovered this point by adjusting
the size parameter of the instantons, which is identified with the baryon size. 
As for the D8 brane profile, we solved it numerically and found that the lowest energy
configuration is well approximated by a simple solution, which is obtained without instantons
for the antipodal configuration. Through our analysis, 
we have restricted to this configuration as a boundary condition of the D8 brane which fixes the dual theory.

In our model, we could find the first order phase transition 
from vacuum to nuclear matter accompanied by a certain jump 
{in}
the baryon density.
{This jump could be interpreted as the effect of the attractive force acting among the baryons
since a finite density of the instanton gas in the DBI action is needed to form a baryonic system. 
Actually, we cannot observe the jump in the models which lack such a term in the DBI action.
While the density is rather large compared to 
{that of}
the nucleus, 
our numerical results obtained near the critical point seems to be reasonable.}
As a bonus, we could find the size of baryon for each $\mu$. It is rather smaller
than the nucleon size just after the nuclear matter is made, but
it increases with the density or chemical potential. This observation
would be reduced to the increasing attractive force among them.


\vspace{.3cm}
{We should notice on the freedom of parameter choice in determining the physical quantities.
For example,}
the numerical values of $n_B$ and $\rho_B$ obtained here depend on our parameter choice.
If we set $M_{KK}\simeq 0.496$ GeV and $\lambda=61.7$ \cite{Seki:2012tt},
then we get
$$
\mu_B\simeq {4.4} \,\mbox{GeV},\quad n_B\simeq 1.4\, \mbox{fm}^{-3},\quad
\rho_B\simeq 0.15\, \mbox{fm}. 
$$
This value of $n_B$ is much 
{larger}
than the ordinary nuclear matter. {In this case, 
we obtain $8\rho_B^3n_B=0.058$, so the nucleons don't still overlap each other in spite of the high density.
Then it is still useful to apply the dilute gas approximation in this parameter choice. Therefore it would be
possible to extend our
analysis given here to high density nuclear matter like neutron stars.}
{Another important extension is to the finite temperature YM theory 
by retaining the chemical potential. Then we could see richer phases of nuclear matter.
This point will be visited in the near future. 
}


\section*{Acknowledgements}
We would like to thank Tetsuo Hatsuda, Shin Nakamura, Kanabu Nawa, Shigeki Sugimoto and
Koichi Yazaki for fruitful comments and discussions. M.T. also would like to thank the hospitality
of Quantum Hadron Physics Laboratory at RIKEN. The work of T.T. is
supported by MEXT/JSPS, Grant-in-Aid for JSPS Fellows No. 24$\cdot$3193.
The work of M.T. is supported in part by the JSPS Grant-in-Aid for Scientific Research, Grant No. 24540280. 
\\

\vspace{2cm}
\noindent{\bf\Large \underline{Appendix}}
\appendix

\section{The properties of the solution $\tau(z)$}\label{app:Simple-sol}
Here we discuss the solutions $\tau(z)$ and show that the antipodal solutions
used here are well approximated by
the simple solution (\ref{simple-tau-sol}). 

\vspace{12pt}

\subsection{Simple solution; antipodal solution for $n=0$; }

The simple solution is obtained
from (\ref{eq-tau-2}) at $n=0$ in the form,
\beq\label{simple-sol}
  \tau_{(0)}(z)=\int_{z_0}^z dz \frac{z_{0}k_0^{5/6}}{zk^{1/6}\sqrt{z^2k^{5/3}-z_{0}^{2}k_0^{5/3}}}\label{eq:sol-tau-n0}\, ,
\eeq
where $k_0=1+z_0^2$ and $z_{0}^{2}k_0^{5/3}=c_0$. This solution has an U-shape form as shown in the 
Fig. \ref{numerical-tau-2}, for example see the one of $z_0=10^{-8}$. 
In the limit of $z_0\to 0$, we find $\tau(\infty)=\pi/2$ and 
that the solution is well approximated by (\ref{simple-tau-sol}). 
Since the energetically favourable solution is obtained at very small $z_0$ as shown below, then 
we find that the solutions used in our analysis
are approximated by the above simple solution which 
has  originally been given in \cite{SS}. In the following, we consider two possible antipodal solutions
for finite $n$.

\vspace{12pt}

\subsection{U-shaped antipodal solutions for finite $n$}
\begin{figure}[htbp]
\begin{center}
\includegraphics[width=8cm]{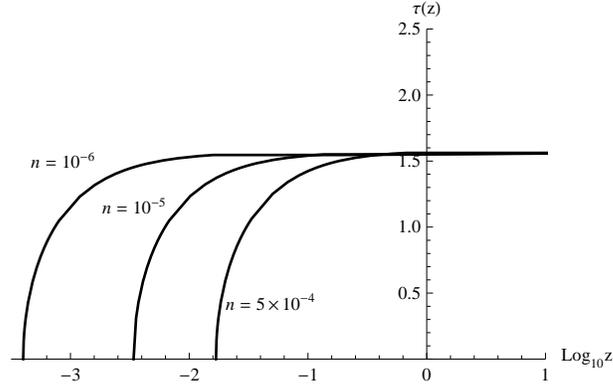}
\caption{{\small Typical examples of the U-shaped solutions of $\tau(\infty)=\pi/2$ for $n=10^{-6}$,
$n=10^{-5}$ and $n=5\times 10^{-5}$.
}}\label{U-shape-tau}
\end{center}
\end{figure}

First of all, we study the solutions for finite $n$ by fixing the value of $\tau(\infty)$ at $\pi/2$.
Namely, we examine the variation of the antipodal solutions when the value of $n$ is increased. 
The numerical results are shown in the Fig. \ref{U-shape-tau}.  
The figure shows that the bottom point $z_0$ of the solution increases with $n$ for the boundary condition of fixed
$\tau(\infty)$ (and $\rho$). 
Then we can use these solutions in our analysis. However we must compare solutions when there are two or more solutions
with the same boundary condition at $z=\infty$. Then we should adopt the one of the lowest
energy density. As shown below, 
the antipodal solutions given above are abandoned in our analysis since their energy densities are larger
than the one of the V-shaped one proposed below. 

\begin{figure}[htbp]
\begin{center}
\includegraphics[width=7cm]{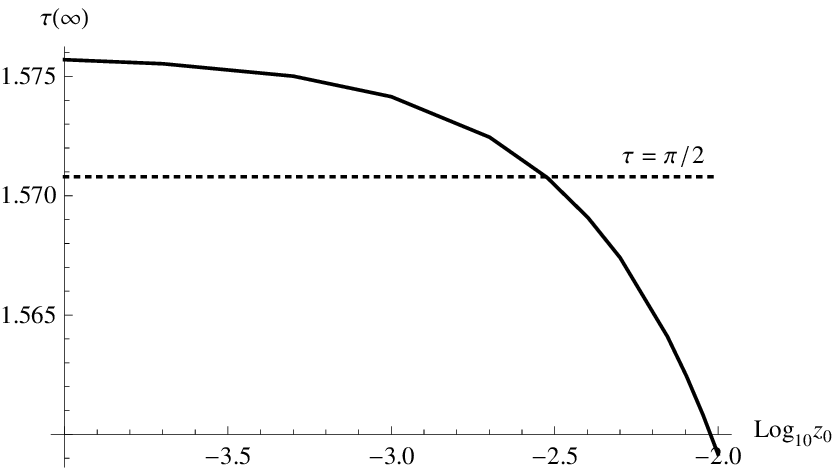}
\includegraphics[width=7cm]{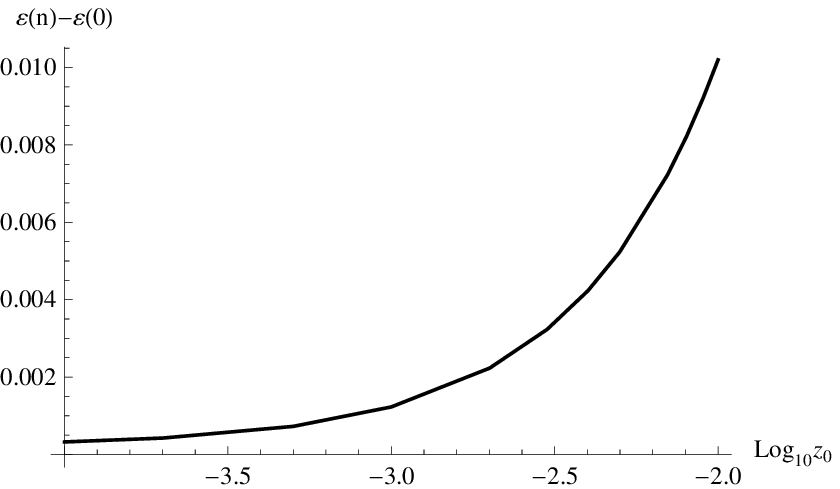}
\caption{{\small (Left) $z_0$ dependence of $\tau(\infty)$. (Right) $z_0$
 dependence of ${\cal E}(n)-{\cal E}(0)$. We take the parameters as
 $n=10^{-5}$ and $\rho=0.05$ in both figures.
}}\label{z0-dependence}
\end{center}
\end{figure}

We show here that the increasing of $z_0$ enlarges the energy density of the U-shaped solutions.
An example of this behavior is shown
in the Fig. \ref{z0-dependence} for $n=10^{-5}$ and
$\rho=0.05$, where we find an antipodal solution at about $z_0\simeq 10^{-2.5}$.
In the right of Fig.~\ref{z0-dependence}, we show the $z_0$ dependence of the energy density
${\cal E}(n)-{\cal E}(0)$ for the corresponding solutions of the left. It increases with $z_0$.\footnote{We
give the definition of ${\cal E}$ in (\ref{energy-d}). In the present case,
we discuss the $n$-dependence of ${\cal E}$ and we fix other parameters
such as $\rho$. We use ${\cal E}(0)$ as a regulator of the energy.} This result implies
that the U-shaped solution with finite $z_0$ would be replaced by smaller energy
antipodal solution with $z_0\to 0$ if it existed. 

\vspace{12pt}
\subsection{Finite $n$ solutions in the limit of $z_0\to 0$}

In order to find an antipodal solution with $z_0\to 0$, we study the solution with $z_0\to 0$ for finite $n$.
This is equivalent to solve the equation with the boundary condition of $z_0\to 0$. In this case, we find
that $\tau(\infty)$ increases with $n$ as shown below.

\vspace{.3cm}
\noindent{\bf $n$ dependence of $\tau(\infty)$ for fixed $z_0$}

In general, the solution of Eq.(\ref{eq-tau-2}) depends on the instanton
density $n$. 
We can see it by
numerical calculation and also by solving the equation expanded
perturbatively by the power series of $n$. 
Especially, we show that $\tau(\infty)$ in the limit $z_0\to 0$ grows 
with increasing $n$.

First, we expand $\tau^\prime$ and $E_z$ as follows
\begin{eqnarray}\label{expand-tau-E}
E_{z} & = &
 E_{z}^{\left(0\right)}+E_{z}^{\left(1\right)}n+E_{z}^{\left(2\right)}n^{2}+
\cdots\,,\\
\tau^{\prime} & = &
 \tau^{\prime\left(0\right)}+\tau^{\prime\left(1\right)}n+\tau^{\prime\left(2\right)}n^{2}+\cdots \,.
\end{eqnarray}
Then, equations of motion of zeroth order of $n$ are given by setting
$n=0$ in the equations (\ref{eq-A0}), (\ref{eq-tau}). 
Imposing boundary conditions as $E_z^{(0)}(z_0)=0$ and $\tau^{\prime
(0)}(z_0)=\infty$, we can solve these equations exactly as follows,
\begin{eqnarray}
E_{z}^{\left(0\right)} & = & 0\label{eq:sol-E-n0}\\
\tau^{\prime\left(0\right)} & = & \frac{z_{0}k_0^{5/6}}{zk^{1/6}\sqrt{z^2k^{5/3}-z_{0}^{2}k_0^{5/3}}}\label{eq:sol-tau-n0-A}\,,
\end{eqnarray}
where $k_0\equiv k(z=z_0)$ for simplicity. Then, we get
$\int_{0}^{\infty}\tau^{\prime(0)}dz=\pi/2$. The solution $\tau^{(0)}$
at $n=0$ is the exact U-shaped solution, which is the same one given above.

Secondly, 
we obtain the next order of equations of motion as follows
\begin{eqnarray}
\partial_{z}\left(k^{5/6}\frac{E_{z}^{\left(1\right)}}{\sqrt{k^{-1/3}+z^{2}\tau^{\prime\left(0\right)2}}}\right) = 6\bar{q}_{0}^{2}\,,\label{eq:eom-E-n1}
\end{eqnarray}
\begin{eqnarray}
\partial_{z}\left\{\frac{\left(z^{2}k^{5/3}-z_{0}^{2}k^{5/3}_{0}\right)^{3/2}}{zk^{3/2}}\tau^{\prime\left(1\right)}
+\frac{3}{2}z_{0}k^{5/6}_{0}\bar{q}_{0}^{2}k^{-1}\left(1-\frac{z^{2}k^{5/3}-z_{0}^{2}k^{5/3}_0}{z^{2}k^{1/3}}\right)\right\}=0 \ .\label{eq:eom-tau-n1}
\end{eqnarray}
We should notice that $\tau^{\prime(1)}$ in Eq.(\ref{eq:eom-tau-n1}) is
independent of $E_z^{(1)}$ because of $E_z^{(0)}=0$. Therefore we can obtain
$\tau^{\prime (1)}$ from Eq.(\ref{eq:eom-tau-n1}) as
\begin{equation}
\tau^{\prime\left(1\right)}\!=\!\frac{zk^{3/2}}{\left(z^{2}k^{5/3}-z_{0}^{2}k^{5/3}_{0}\right)^{3/2}}\!\!\left(\!C_{\tau}^{\left(1\right)}\!-\!\frac{3}{2}\frac{z_{0}k^{5/6}_{0}\bar{q}_{0}^{2}}{k}\!\!\left(\!1\!-\!\frac{z^{2}k^{5/3}-z_{0}^{2}k^{5/3}_{0}}{k^{1/3}z^{2}}\!\right)\!\right),\label{eq:sol-tau1}
\end{equation}
where $C_{\tau}^{\left(1\right)}$ is an integration constant. This
expression seems to be divergent as ${\cal O}(\epsilon^{-3/2})$ when $z$ is
close to $z_0$ as $z=z_0+\epsilon$, and the degree of divergence is
larger than that of $\tau^{\prime (0)}(z_0)$. Thus, we need to choose the
integration constant as
\begin{equation}
C_{\tau}^{\left(1\right)}=\frac{3}{2}z_{0}k_{0}^{-1/6}\bar{q}_{0}^{2}\left(z_{0}\right)\,.\label{eq:Ctau-n1}
\end{equation}
Then the divergence becomes
same as that of $\tau^{\prime (0)}(z_0)$. Since the value of $\tau^{\prime (1)}$ is
always positive in any $z$ when $z_0=0$, we obtain
\begin{eqnarray}
\tau(\infty)\simeq \int^{\infty}_{0}\left(\tau^{\prime (0)}
+\tau^{\prime (1)}n  \right)dz>\frac{\pi}{2}\,.
\end{eqnarray}
Therefore, $\tau(\infty)$ is always larger than $\pi/2$ at finite $n$.
\begin{figure}[htbp]
\begin{center}
\includegraphics[width=8cm]{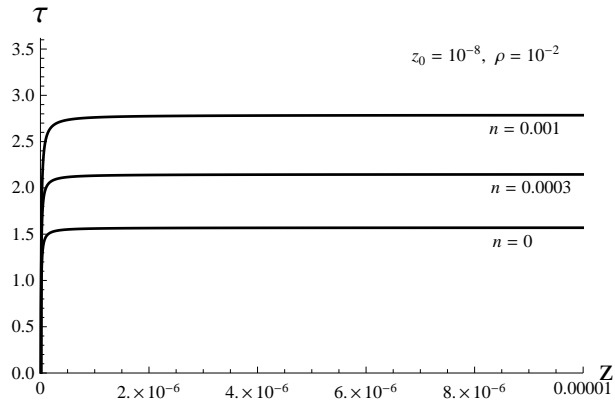}
\caption{{\small Typical examples of the solutions of $\tau(z)$ for $z_0=10^{-8}$
 and $\rho=10^{-2}$. 
}}\label{numerical-tau-2}
\end{center}
\end{figure}
We can ensure the $n$-dependence of $\tau(\infty)$ by numerical solutions.
Typical examples are shown at small $z_0=10^{-8}$ for finite $n$ in Fig.~\ref{numerical-tau-2}. We assure that
$\tau(\infty)$ increases with $n$.

\vspace{12pt}

\noindent{\bf V-shaped antipodal solutions}

From the above result, the solution with $z_0\to 0$ at finite value of $n$ is not the antipodal solution. 
However, we can construct an antipodal solution by using this solution according to the idea of \cite{BLL}.
First, cut off the part of $z_0<z<z_c$ of the above solution for $D8$, then connect it with the corresponding
cut one of $\overline{D8}$. This connected configuration has a V shape, and the point $z_c$ is chosen such that
the the distance between two end points of $D8$ and  $\overline{D8}$ becomes $\pi$, namely the antipodal
configuration.

In general, however, the V-shaped solution has a cusp at the connected point $z=z_c(\geq z_0)$ because
of $\tau'(z_c)\neq \infty$, and they
need a source term \footnote{For example, we can consider the $D_4$ brane as in  \cite{BLL}, however it is not
necessary here to specify it.} in order to cancel this cusp singularity. Then, the total
action should be replaced as
\beq
S=S_{DBI}+S_{CS}\quad \Longrightarrow \quad \tilde{S}=S+S_{source}\,.
\eeq
As we can see from the Fig.~\ref{numerical-tau-2}, the value of $z_c$ is estimated
as $z_c\simeq z_0$. Then, the V-shaped solution is almost the same one
with the simple solution given above. In addition, the $z$-component of the tension of
the D8 brane on the cusp is negligible small because of $z'(\tau)\simeq 0$ at
the $z=z_c$, then the contribution of the source for the force balance
conditions is negligible. Therefore, we can ignore the source term.

\begin{figure}[htbp]
\begin{center}
\includegraphics[width=8cm]{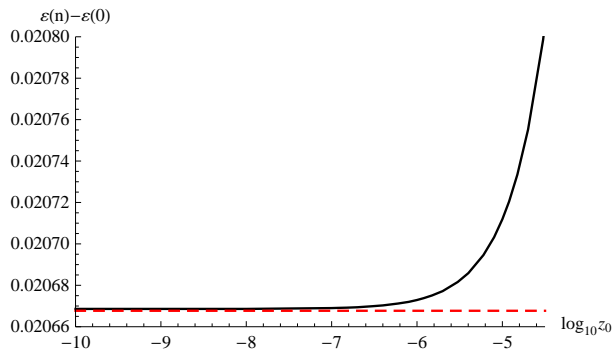}
\caption{{\small V-shaped solution v.s. simple solution for $n=10^{-3}$
 and $\rho=10^{-2}$. The curve denotes the $z_0$
 dependence of ${\cal E}(n)-{\cal E}(0)$ of the V-shaped one. The
 dotted (red) line denotes the value of ${\cal E}(n)-{\cal E}(0)$
 of the simple solution. 
}}\label{simple-v-shape}
\end{center}
\end{figure}

The next problem is to determine which antipodal solution we adopt in the present analysis. 
This is performed by comparing the energy densities of the two kinds of solution. As the example,
we have compared the energy density (${\cal E} (U,n)$) of the three examples given in the Fig. \ref{U-shape-tau}
with the V-shaped (${\cal E}(V,n)$) one given here, then we find ${\cal E}(U)/{\cal E}(V)\sim 20$. As shown in this
example, the answer is that we should
use the V-shaped one with $z_0\to 0$ since its energy density is the lower than that of U-shaped one
for finite $n$. 
And the adopted V-shaped solution is well approximated by our simple solution since the energy densities
of them are almost equal as shown in the Fig. \ref{simple-v-shape}. 
Thus, the simple antipodal solution (\ref{simple-tau-sol}) is useful in our analyses for any value of $n$.

\begin{thebibliography}{99}
 
 \bibitem{Schafer:2005ff}
  T.~Schafer,
  ``Phases of QCD,''
  hep-ph/0509068.

 
 \bibitem{Stephanov:2007fk}
  M.~A.~Stephanov,
  ``QCD phase diagram: An Overview,''
  PoS LAT {\bf 2006}, 024 (2006)
  [hep-lat/0701002].

   
 \bibitem{SS}
   T.~Sakai and S.~Sugimoto,
   ``Low energy hadron physics in holographic QCD,''
   Prog.\ Theor.\ Phys.\  {\bf 113}, 843 (2005)
   [hep-th/0412141].   
 \bibitem{HSSY}
   H.~Hata, T.~Sakai, S.~Sugimoto and S.~Yamato,
   ``Baryons from instantons in holographic QCD,''
   Prog.\ Theor.\ Phys.\  {\bf 117}, 1157 (2007)   [hep-th/0701280 [HEP-TH]].

{
 \bibitem{Hashimoto:2008zw} 
   K.~Hashimoto, T.~Sakai and S.~Sugimoto,
   ``Holographic Baryons: Static Properties and Form Factors from Gauge/String Duality,''
  Prog.\ Theor.\ Phys.\  {\bf 120}, 1093 (2008)
  [arXiv:0806.3122 [hep-th]].

\bibitem{Hashimoto:2009ys} 
  K.~Hashimoto, T.~Sakai and S.~Sugimoto,
  ``Nuclear Force from String Theory,''
  Prog.\ Theor.\ Phys.\  {\bf 122}, 427 (2009)
  [arXiv:0901.4449 [hep-th]].


\bibitem{Kim:2006gp} 
  K.~-Y.~Kim, S.~-J.~Sin and I.~Zahed,
  ``Dense hadronic matter in holographic QCD,''
  hep-th/0608046.

\bibitem{BLL} 
  O.~Bergman, G.~Lifschytz and M.~Lippert,
  ``Holographic Nuclear Physics,''
  JHEP {\bf 0711}, 056 (2007)
  [arXiv:0708.0326 [hep-th]].

\bibitem{Kim:2007vd} 
  K.~-Y.~Kim, S.~-J.~Sin and I.~Zahed,
  ``Dense holographic QCD in the Wigner-Seitz approximation,''
  JHEP {\bf 0809}, 001 (2008)
  [arXiv:0712.1582 [hep-th]].

\bibitem{RSRW} 
  M.~Rozali, H.~-H.~Shieh, M.~Van Raamsdonk and J.~Wu,
  ``Cold Nuclear Matter In Holographic QCD,''
  JHEP {\bf 0801}, 053 (2008)
  [arXiv:0708.1322 [hep-th]].

\bibitem{Chuang:2010ku} 
  W.~-y.~Chuang, S.~-H.~Dai, S.~Kawamoto, F.~-L.~Lin and C.~-P.~Yeh,
  ``Dynamical Instability of Holographic QCD at Finite Density,''
  Phys.\ Rev.\ D {\bf 83}, 106003 (2011)
  [arXiv:1004.0162 [hep-th]].

\bibitem{Kaplunovsky:2012gb} 
  V.~Kaplunovsky, D.~Melnikov and J.~Sonnenschein,
  ``Baryonic Popcorn,''
  JHEP {\bf 1211}, 047 (2012)
  [arXiv:1201.1331 [hep-th]].

\bibitem{Seki:2012tt} 
  S.~Seki and S.~-J.~Sin,
  ``Chiral Condensate in Holographic QCD with Baryon Density,''
  JHEP {\bf 1208}, 009 (2012)
  [arXiv:1206.5897 [hep-th]].

\bibitem{deBoer:2012ij} 
  J.~de Boer, B.~D.~Chowdhury, M.~P.~Heller and J.~Jankowski,
  ``Towards a holographic realization of the Quarkyonic phase,''
  arXiv:1209.5915 [hep-th].


}

\bibitem{Cohen2004} 
  T.~D.~Cohen,
   ``QCD functional integrals for systems with nonzero chemical potential,''
  In *Shifman, M. (ed.) et al.: From fields to strings, vol. 1* 101-120
  [hep-ph/0405043].

\bibitem{HPY}
  P.~Haensel, A.~Y.~Potekhin and D.~G.~Yakovlev,
  ``Neutron stars 1: Equation of state and structure,''
  (Astrophysics and space science library. 326)

\end{thebibliography}

\newpage

\end{document}